\documentclass[sigconf]{acmart}

\setcopyright{none}
\renewcommand\footnotetextcopyrightpermission[1]{}
\usepackage{amsmath}
\usepackage{amsfonts}
\usepackage{booktabs}

\AtBeginDocument{%
  }

\acmConference[RecSys CONSEQUENCES '26]{ACM Conference on Recommender Systems}{October 2, 2026}{Minneapolis, MN, USA}
\acmYear{2026}
\copyrightyear{2026}

\begin{document}

\title{Beyond Raw Engagement: A Counterfactual Observability Framework for Recommender Systems at Netflix}

\author{Chaoran Guo}
\authornote{These authors contributed equally.}
\orcid{0000-0002-0877-7063}
\email{cguo@netflix.com}
\affiliation{\institution{Netflix}\city{Los Gatos}\state{California}\country{USA}}

\author{Ding Tong}
\authornotemark[1]
\orcid{0000-0003-3148-280X}
\email{dingt@netflix.com}
\affiliation{\institution{Netflix}\city{Los Gatos}\state{California}\country{USA}}

\author{Ting-Po Lee}
\authornotemark[1]
\authornote{Work done at Netflix}
\orcid{0000-0002-9421-8566}
\email{tingpo.lee@gmail.com}
\affiliation{\city{San Jose}\state{California}\country{USA}}

\author{Scarlet Chen}
\email{sijiac@netflix.com}
\affiliation{\institution{Netflix}\city{Los Gatos}\state{California}\country{USA}}

\renewcommand{\shortauthors}{Guo, Tong, Lee, and Chen.}

\begin{abstract}
Understanding the performance of large-scale recommender systems remains an underexplored challenge, especially for content creators and model developers. The raw engagement signals available to them, such as views and clicks, conflate content quality, model behavior, presentation bias, and audience reach, making it hard to attribute outcomes to the right cause.

In this work, we present a general evaluation framework that enhances observability across multiple recommender systems at Netflix and demonstrate its effectiveness through several production deployments. The framework treats recommender-system observability as a counterfactual measurement problem: estimating what the recommender would have done, and what engagement would have followed, in the absence of a specific content item or model decision. We articulate three stakeholder-centered observability principles for content creators and model developers, and propose measurement methodologies covering bias reduction, relativity, and incrementality, applicable to both single-stage and cascading recommender systems and serving both audiences from a single measurement foundation. 

\end{abstract}



\maketitle

\section{Introduction}
Recommender systems power personalized experiences across media, e-commerce, advertising, and information retrieval, yet most research optimizes user engagement while overlooking the content creators and model developers who need to understand \emph{why} the system behaves as it does. Neglecting these stakeholders can harm a platform's long-term health~\cite{creator,creator2,su2024explore}. The performance signals available to them are typically raw statistics such as views, clicks, and likes, which are heavily biased by the recommendation and presentation process~\cite{biassurvey}; without scientific measurement and bias correction, they receive misleading signals about what works and why. For example, a high-quality item can appear to underperform on click-through rate (CTR) simply because a newer model positions it poorly, leading to undervaluation of good content or misguided optimization.

We present an evaluation framework, designed and deployed at Netflix over several years, that improves observability across large-scale recommender systems. Its central idea is to treat observability as a \emph{counterfactual measurement problem} over the full pipeline: estimating what the recommender would have done, and what engagement would have followed, in the absence of a specific content item or model decision. This reframing is what lets us separate content quality from model behavior and deliver bias-corrected signals to both content creators and model developers. Our contributions are: 
\begin{itemize}
    \item \textbf{Stakeholder-centered observability principles.} Drawing on years of evaluation experience across recommender systems at Netflix, we articulate three principles for recommender-system observability. These principles target the needs of both content creators and model developers.

    \item \textbf{A unified measurement toolkit.} Starting from an Exploration \& Exploitation module with logged selection probabilities, we develop measurements for bias reduction, relativity, and incrementality. In particular, we propose incrementality measurements for both single-stage and cascading recommender systems under a unified framework. Together, these measurements uphold the observability principles by decoupling content quality from model behavior and correcting for recommendation-induced bias, addressing two central challenges in production recommender-system observability.

    \item \textbf{Production validation and deployment.} We validate the framework through simulations against ground truth, strong alignment with historical A/B tests at Netflix, and a production monitoring case study where incrementality metrics helped surface a recommender-system label misattribution issue, illustrating how the framework supports the long-term health of the recommender system.
\end{itemize}
Beyond the Netflix deployment, our measurement framework are formulated generically and apply to other recommender systems with exploration-based logging, offering a practical observability template for the broader RecSys community.

\section{Common Observability Pitfalls}
At Netflix, we have been developing evaluation methodologies for various recommender systems over the years, and we noticed a few common pitfalls in the industry in building the observability of content creators and model developers.

\textbf{Limited transparency of the ML ``black box''.} Many production recommenders expose only high-level logs or aggregate counters. This lack of diagnostic data leaves both content creators and model developers guessing when performance drops.

\textbf{Biased evaluation signals.}
\textit{Presentation bias}. Higher-ranked content receives more attention regardless of quality, inflating click and view counts.
\textit{Content performance confounded with model performance}. Content creators may draw incorrect conclusions if only relying on raw engagement metrics such as Click-Through Rate (CTR), which often reflect both the quality of the content and the effectiveness of the model's recommendations. For example, a recommended content item might seem to underperform with low CTR simply because a newer model is not positioning it effectively, not due to lower quality; relying on CTR can lead to undervaluation of high-quality content or misguided optimization efforts.

\textbf{Short-horizon objectives.} Many evaluation frameworks optimize for immediate interactions, which can discourage exploration and hurt long-term outcomes such as user retention or content diversity.

To address these pitfalls, we propose three \textbf{principles} for an effective evaluation framework: 

\begin{itemize}
    \item Separate content quality from model behavior.
    \item Correct for different types of bias that appear in recommender systems.
    \item Align measurements with long-term business value. 
\end{itemize}
These principles guided the design of our evaluation framework introduced in the next section.

\section{Evaluation Methodology}
\subsection{Measurement Overview}
Inspired by recent advancements in causal inference, we propose a suite of metrics focusing on bias reduction, relativity, and incrementality \cite{Sato_2020} to measure the performance of recommended content, agnostic to model exposure.

We consider a contextual bandit or reinforcement learning (RL) setup for the Exploration \& Exploitation module in our recommender systems (see Appendix \ref{system-design} for details). Each arm $a_j$ is selected following policy $\pi$ with probability $\pi(a_j \mid x_j)$ given context $x_j$. If $a_j$ is chosen, we observe reward $r_j$, which can be a short-term reward like click-through rate (CTR) or a long-term reward like user retention.

For \textbf{bias reduction}, we consider the inverse propensity score (IPS) \cite{pmlr-v32-agarwalb14} and its clipped variations \cite{clip}, to correct for over- or under-representations of content. For the target policy $\pi_e$ under logged policy $\pi_0$, the self-normalized IPS (SNIPS) over $N$ logged samples is computed as:
\begin{equation*}
   \text{SNIPS} =
   \frac{\sum_{j=1}^{N} r_j
   \frac{m(a_{j,\pi_e}, a_{j,\pi_0})}{\pi_0(a_j \mid x_j)}}
   {\sum_{j=1}^{N}
   \frac{m(a_{j,\pi_e}, a_{j,\pi_0})}{\pi_0(a_j \mid x_j)}} ,
\end{equation*}
where $m$ is the match function\footnote{Exact matching is a special case that provides unbiased estimates. However, other matching functions can be employed for various bias-variance tradeoffs.} that evaluates the degree of alignment between the target policy and logged policy. 

For \textbf{relativity}, we ask how one entity compares against another. This applies in two ways: at the policy level (one algorithm versus another) and at the item level (one content item versus others). Together, these two components let us separate algorithm performance from content performance.

For \emph{policy-level relativity}, we compare two policies by computing SNIPS with each as the target policy and taking the difference. 

For \emph{item-level relativity}, we ask: ``how does this content item perform against other content shown at the same position?'' We use a position-adjusted SNIPS. Assuming a content set $\bar{c}$ containing $|\bar{c}|$ pieces of content eligible at $K$ positions is reviewed by users, we first compute the difference between content $c_i$ at position $k$ and a random content from $\bar{c}$ at the same position, to mitigate presentation bias. Since content may appear at different positions, we aggregate the lift across all $K$ positions for $c_i$, weighted by the position-adjusted weight $w_k$ at each position $k$: $\sum_{k=1}^{K} w_{k} (\text{SNIPS}_{c_{i},k} - \text{SNIPS}_{\bar{c}, k})$.

\textbf{Incrementality.} Most importantly, for content we ask: \emph{if this item were removed, how much value would we lose once the recommender substitutes its next-best alternative?} This counterfactual view reveals an item's true contribution and is one of the keys to separating intrinsic quality from model behavior. We detail incrementality for single-stage and cascading systems next.

\subsection{Incrementality Measurement for Single-Stage Recommender System}
\label{sec:loo-single}
A single-stage (non-cascading) recommender decides in one layer, without passing candidates through upstream stages. We propose a \emph{Leave-One-Out} measurement: a simulated policy, built on top of the Exploration policy, for the counterfactual in which a specific item is removed from the content set. Using the SNIPS of the Exploration policy and the Leave-One-Out policy, the incrementality is the lift aggregated across positions,
\begin{equation}
\sum_{k=1}^{K} w_{k}\,(\text{SNIPS}_{\text{Exploration},k} - \text{SNIPS}_{\text{Leave-One-Out},k}).
\end{equation}
The match function $m=1$ if (i) the explored arm is the Exploration policy's top pick and is not the removed content, or (ii) it is the second-best pick and the top pick is the removed content; otherwise $m=0$. To reduce variance, we cap logged probabilities at the 10th percentile as a lower bound.

\subsection{Incrementality Measurement in Cascading Recommender Systems}
\label{loo_set_lift}
In cascading recommender systems, each stage derives its candidate pool from upstream stages (e.g., ranking operates on retrieved candidates, reranking on the ranked set). A content item's observed performance thus depends on every stage, and a low engagement signal may reflect upstream filtering as much as intrinsic quality. The right measurement depends on the audience.

\emph{Upstream model developers} ask whether including item $i$ in a candidate pool was a good decision. We reuse the Leave-One-Out construction, but because dropping $i$ changes the entire downstream pool, the target action becomes a set rather than a single top pick; we therefore redefine the match $m$ as a \emph{set similarity} (e.g., Jaccard) between the target and explored candidate pools, and aggregate the per-item lift with the same formula as the single-stage case (Appendix~\ref{app:estimators}). High lifts across many items indicate the upstream model contributes genuine value; consistently low lifts mean its pools are interchangeable downstream.

\emph{Content creators and platform observability} instead need an item's overall contribution across the whole cascade, accounting for both how often it is eligible and how much value it adds when shown. Because a Leave-One-Out policy is expensive to reproduce faithfully across a full pipeline, we decompose the overall value into two interpretable quantities:
\begin{equation}
\text{Incrementality}(i) = \text{Irreplaceability}(i)\cdot\text{Universality}(i),
\end{equation}
where \emph{Irreplaceability} is the causal incremental lift conditional on $i$ being eligible in its candidate pool, and \emph{Universality} is the probability that $i$ is eligible. We estimate Irreplaceability from \textbf{synthetic treatment samples} built on the Exploration module's logged selection probabilities, using a self-normalized H\'ajek IPS estimator; the Horvitz--Thompson form, the without-replacement sequential-generation variant, and the Universality estimator are given in Appendix~\ref{app:estimators}. This decomposition separates conditional content quality from system reach, which is essential for comparing niche content against broadly appealing content.

\section{Results}
Our evaluation framework has been widely adopted across recommender systems at Netflix. Given the breadth of deployment, we present here only a few representative case studies that validate the accuracy of our methodology, including a simulation validating the Leave-One-Out approach against the simulated truth (Section~\ref{sec:single-stage-results}), an online A/B test alignment study (Section~\ref{sec:ab-validation}), and a model developer observability case study (Section~\ref{sec:model-dev-case}).

\subsection{Results for Single-Stage Recommender System} \label{sec:single-stage-results}
We perform simulations to verify the accuracy of our measurement methodology, focusing on the incrementality measurement to capture the true performance of content items, independent of specific models. 

In the simulation, we start with a set of content items, emulate both a target policy and an Exploration policy that recommend the content items to users, and then get the click-through rate (CTR) of these items $CTR_{FullContentSet}$. We randomly pick two items from the content items, denoted as R1 and R2. Then, we simulate the scenarios where R1 or R2 is removed from the content items, and get the corresponding click-through rate $CTR_{R_1 Removed}$ and $CTR_{R_2 Removed}$. In this way, the true incrementality of R1 and R2 will be the $CTR_{FullContentSet} - CTR_{R_1Removed}$ and $CTR_{FullContentSet} - CTR_{R_2Removed}$ respectively, which can be then compared against our Leave-One-Out measurement as described in Section~\ref{loo_set_lift} to validate our measurement methodology. We also apply different noise levels to the simulations. The detailed setup is described in the Appendix.

\begin{figure*}[htbp]
    \centering
    
    \begin{minipage}{0.45\textwidth}
        \centering
        \includegraphics[width=\linewidth]{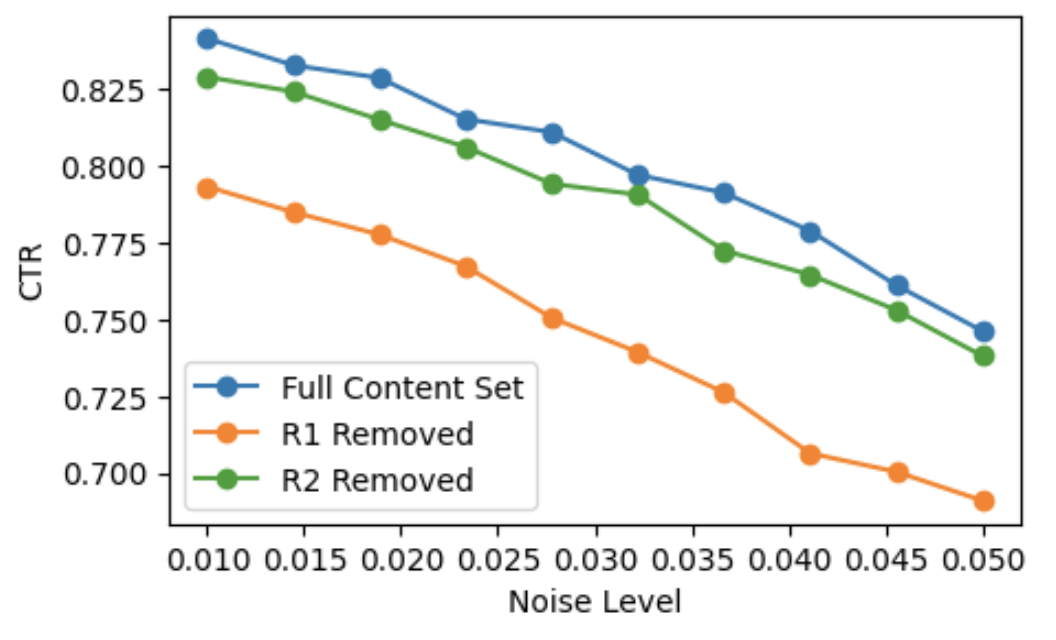}
        \vspace{0.1cm}
        \\ (a) True CTR ($CTR_{FullContentSet}$, $CTR_{R_1Removed}$, $CTR_{R_2Removed}$) in the simulation.
        \label{fig:simu_page} 
    \end{minipage}
    \hfill
    \begin{minipage}{0.45\textwidth}
        \centering
        \includegraphics[width=\linewidth]{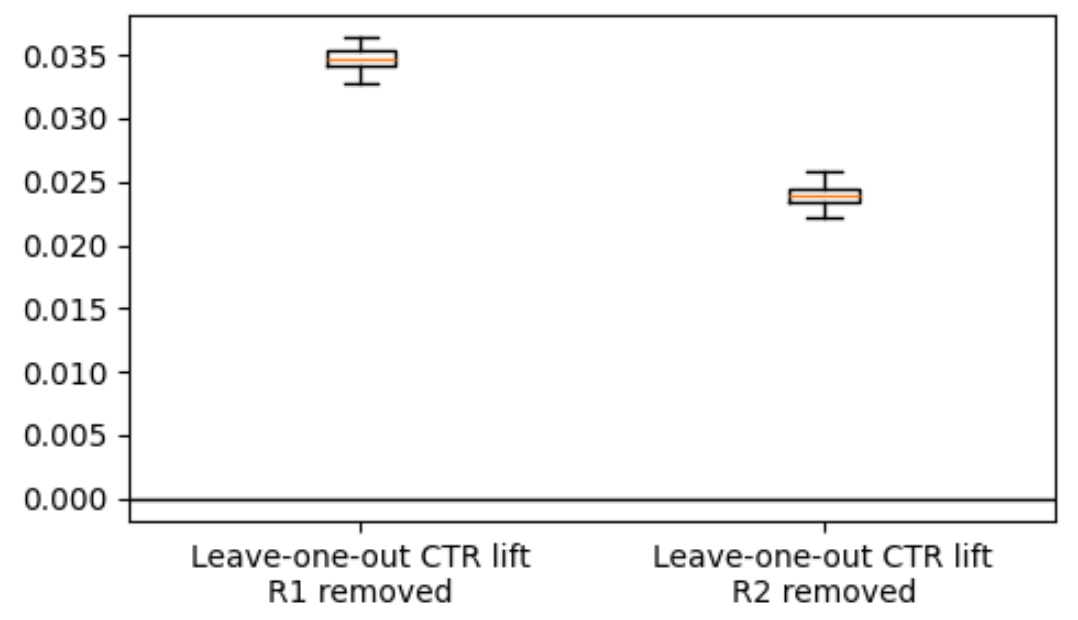}
        \vspace{0.1cm}
        \\ (b) Leave-One-Out content set CTR lift of R1 and R2
        \label{fig:cee_lift} 
    \end{minipage}
    
    \caption{Leave-One-Out CTR lift aligns with the simulated CTR lift. Left (a): Simulation results. Right (b): Leave-One-Out lift.}
    \label{fig:simulated_vs_lift} 
\end{figure*}

The results are shown in Figure~\ref{fig:simulated_vs_lift}. The true incrementality of R1 and R2 can be seen from the left panel (a), where the blue line shows $CTR_{FullContentSet}$ when all content are included, the orange line represents the removal of R1, and the green line shows the removal of R2. The incrementality value of R1 and R2 will be their distances from the blue line respectively. Our Leave-One-Out measurement is shown in the right panel (b) for content item R1 and R2. Comparing the two plots, we find that the incremental value of R1 and R2 align closely between these two methods. The magnitude difference is very small at small noise level of the simulation, and becomes larger as the noise goes up, but still tolerable. This finding validates the use of our metrics in evaluating the incremental value of individual content items. 


\subsection{Alignment with Online A/B Testing Results} \label{sec:ab-validation}
To validate that our approach works for content creators, we conducted online A/B tests in which content items were removed from the recommender system, allowing us to compare our incrementality measurements against source-of-truth results from the tests.

Here we introduce a case study for Row Curators. Rows are collections of titles on Netflix homepages curated by content creators, who want to speed up the curation by understanding the incremental impact of each curation offline without running A/B tests. Here we obtain ground truth values by removing series of rows from the recommender system and measuring the resulting loss in user engagement, then compare our proposed measurement against this A/B-test ground truth.

For the Irreplaceability estimate, we tried different estimators and eventually selected the self-normalized H\'ajek estimator for robust observability signal under such heterogeneous propensities. In practice, we clip the propensities to $[0.0001, 0.9999]$ to further reduce variance, require at least 100 observations for each row, and aggregate across homepages of varying lengths. Across 11 A/B tests, the $R^2$ between our proposed row-incrementality measurement and the A/B-test ground truth is $0.8212$, providing strong validation of our approach. 

Given this high degree of alignment, row curators can leverage our proposed row incrementality measurement offline to learn which curation ideas and strategies work better, devise stronger curation strategies, and drive positive user impact.

\begin{figure}[t]
  \centering
  \includegraphics[width=\linewidth]{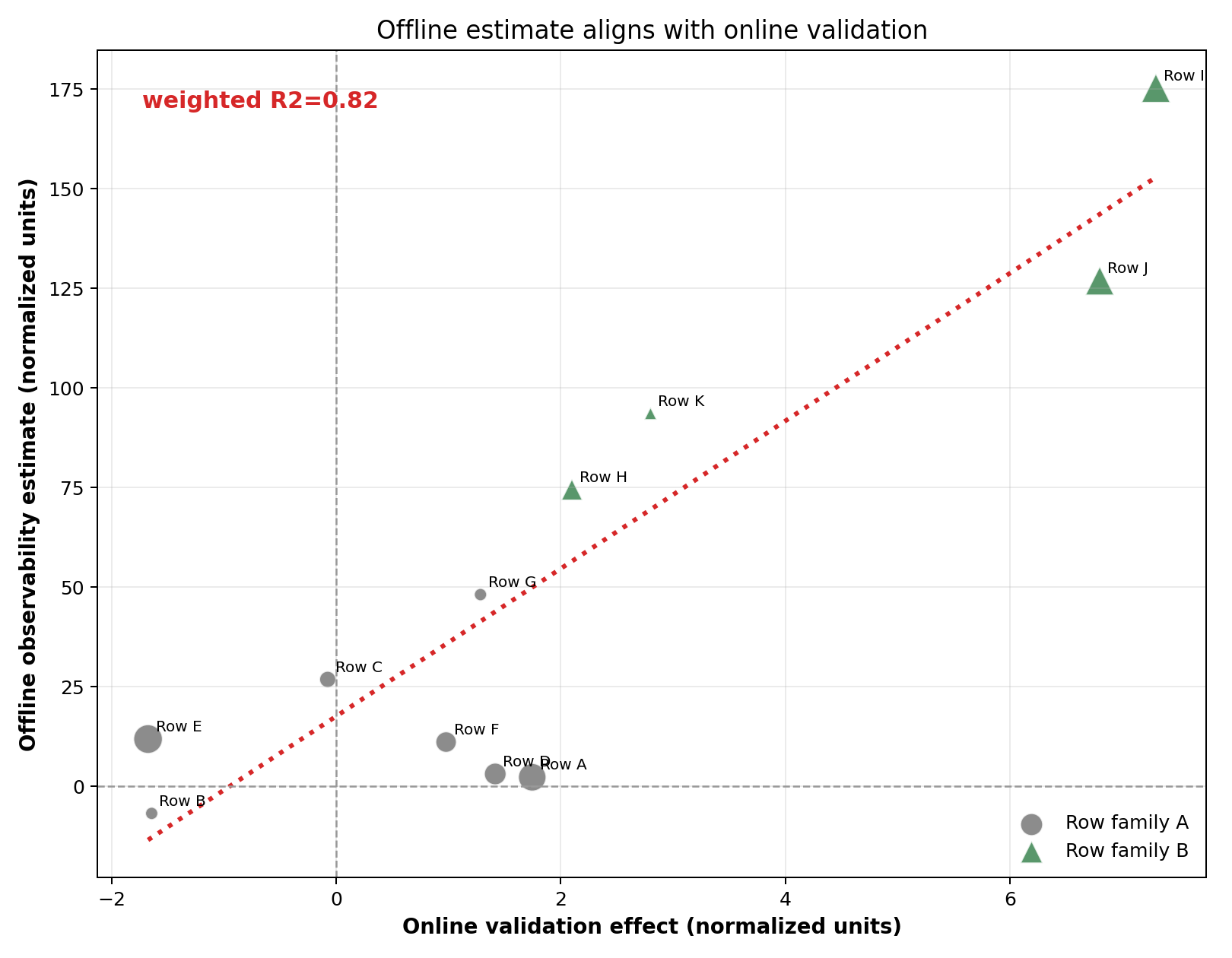}
  \caption{Validation of offline row incrementality against historical A/B-test user engagement effects. Point size is proportional to the number of observation samples we have; marker shape separates two different family of homepage rows.}
  \Description{A scatter plot compares historical A/B-test user engagement effects with proposed offline row incrementality measurement for 11 homepage rows.}
  \label{fig:ab-validation}
\end{figure}

\subsection{Model Developer Observability Case Study} \label{sec:model-dev-case}
We observed multiple successful cases after deploying our evaluation framework across different recommender systems. In one case, we built a monitoring mechanism to track content performance metrics over time and spot anomalies. This helped uncover gaps in our recommenders and enabled model developers to quickly identify root causes and improve model quality.

\begin{figure}[t]
  \centering
  \includegraphics[width=\linewidth]{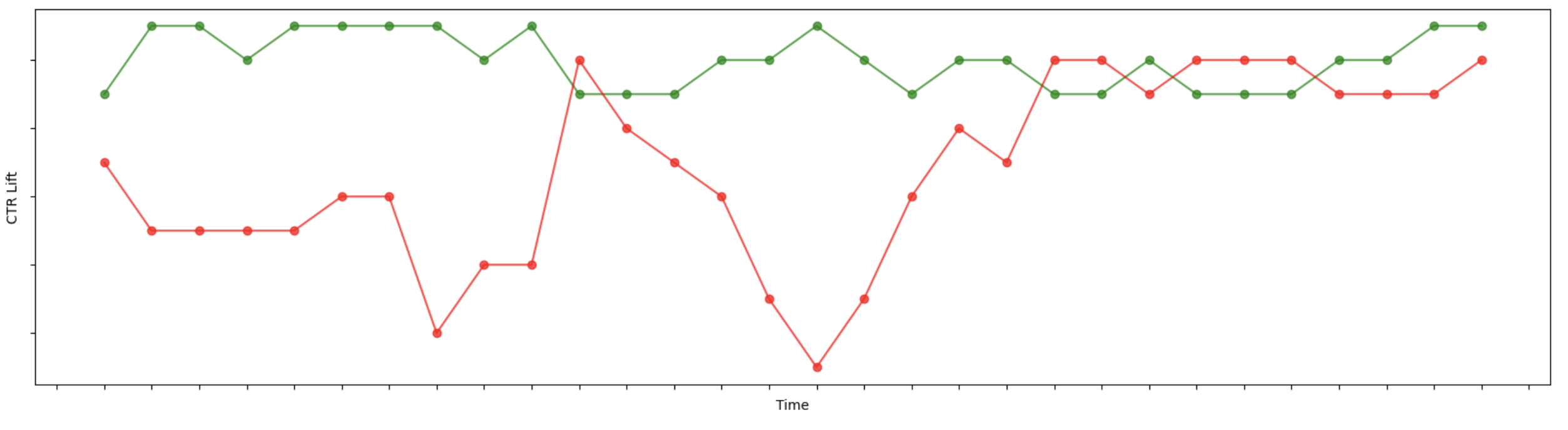}
  \caption{Monitoring time series of content incrementality metrics.}
  \label{fig:time-series}
\end{figure}

In Figure~\ref{fig:time-series}, we show how we detected a sudden dip in the performance of content A, marked in red, while similar content remained relatively stable. This unexpected dip led to the discovery of a label misattribution issue affecting content A's positive labels in our model. This example highlights the role of the evaluation framework in monitoring long-term system health and surfacing actionable insights to model developers.

\section{Conclusion}
  This paper presents a generic evaluation framework that elevates the observability in large-scale recommender systems: we estimate what the recommender would have done, and what engagement would have followed, in the absence of a specific content item or model decision.The framework rests on three stakeholder-centered observability principles and is operationalized through a suite of measurement for bias reduction, relativity, and incrementality, with applications spanning single-stage and cascading recommender systems in a generic way. A key design choice is that the same measurement foundation serves two audiences with distinct needs: content creators, who need interpretable per-item signals to guide their strategic decisions, and model developers, who need diagnostic signals to assess and improve their recommender systems. We validate the framework through simulations against ground truth, alignment with online A/B tests at Netflix, and a production monitoring case
  study, demonstrating practical value at deployment scale. A useful operational consequence of these validations is that many measurement questions can now be answered offline, making iterations much faster.

  By integrating these measurements into the production feedback loop, the generic framework fosters a positive flywheel effect for multiple recommender systems: better measurements lead to higher-quality items generated from content creators and sharper decisions from model developers; the resulting improvements attract more users and generate richer signals for the recommender; and those signals in turn enable even better measurements. Observability, treated this way, becomes not a downstream report but a generative loop that compounds over time, benefiting content creators, model developers, and the broader recommendation ecosystem.

\bibliographystyle{ACM-Reference-Format}
\bibliography{references}

\section{Appendix}
\subsection{System Design}\label{system-design}

\begin{figure}[htbp]
    \centering
    
    \begin{minipage}{0.8\linewidth}
        \centering
        \includegraphics[width=\linewidth]{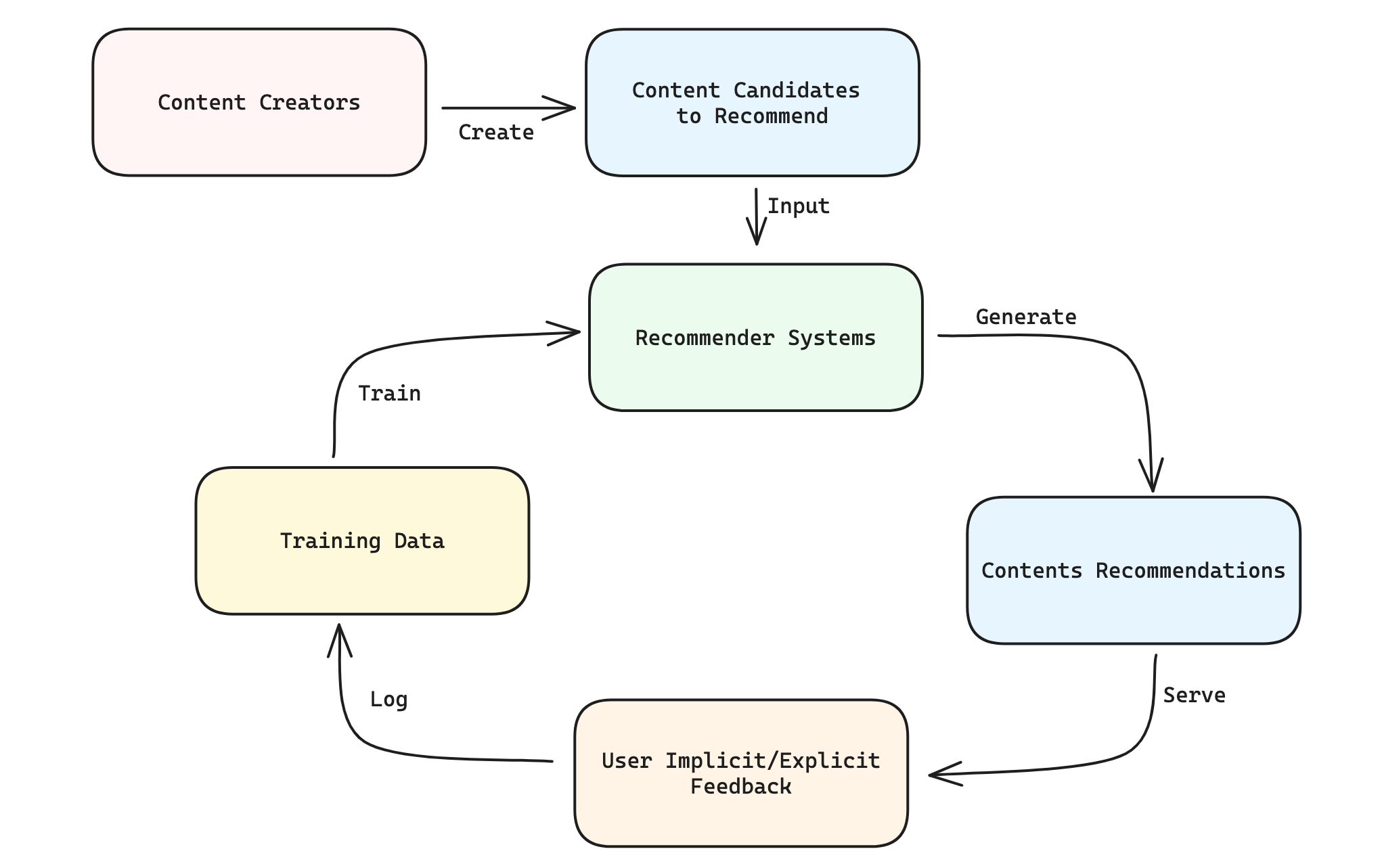}
        \vspace{0.1cm}
        \\ (a) Traditional Recommender System
    \end{minipage}
    
    \vspace{0.4cm} 
    
    \begin{minipage}{0.8\linewidth}
        \centering
        \includegraphics[width=\linewidth]{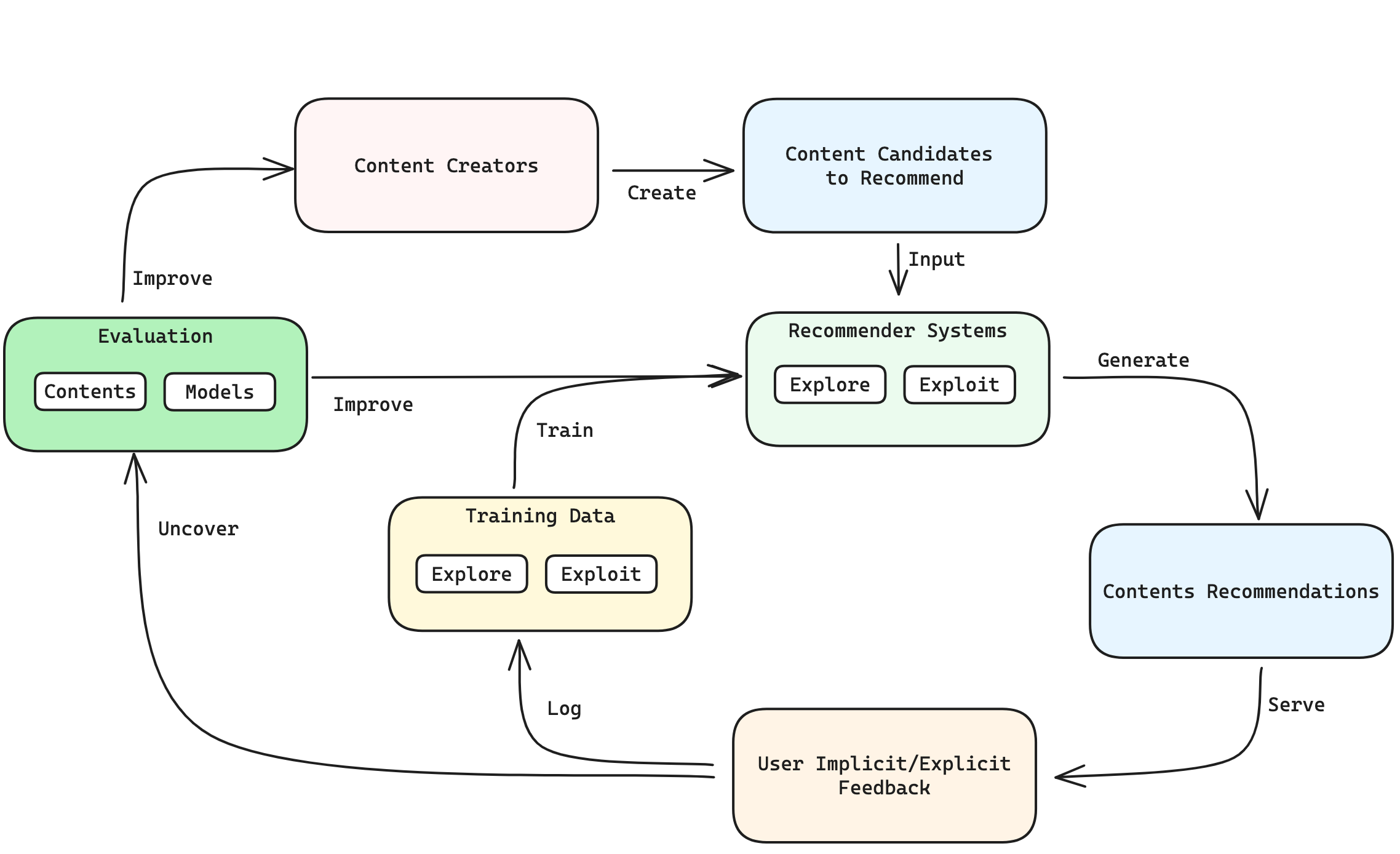}
        \vspace{0.1cm}
        \\ (b) Recommender System with Enhanced Observability
    \end{minipage}
    
    \caption{Recommender System Design to Improve Observability. Top (a): Traditional approach. Bottom (b): Enhanced approach.}
    \label{fig:main_system}
\end{figure}

Many existing recommender systems often leave the observability of content creators and model developers outside of the model training loop (Figure~\ref{fig:main_system}(a)). In contrast, our proposed evaluation system (Figure~\ref{fig:main_system}(b)) leverages user feedback data to develop a measurement framework that \textbf{separates intrinsic content performance from the effectiveness of the recommender system}, starting with adding an optimized Exploration \& Exploitation module. The Exploration \& Exploitation module not only provides the foundation to debias measurements and decouple content performance from model performance, but also enables us to optimize for long-term user value on the platform \cite{su2024explore}. We then measure recommended content performance using the methodology proposed below, and the insights are continuously integrated back into the ecosystem to create a flywheel of data collection, evaluation, content creation, model refinement, and improved content delivery.

\subsection{Estimator Details}
\label{app:estimators}

\subsubsection{Upstream Set-Similarity Leave-One-Out}
In a cascade, dropping item $i$ at the upstream stage affects the entire candidate pool that flows downstream, so the target action is a set rather than a single top pick, and exact matching is almost never achieved (any missing or reordered item makes $m=0$). We therefore keep the Leave-One-Out policy construction unchanged but redefine $m$ as a \emph{set similarity} (e.g., Jaccard) between the target policy's candidate pool and the explored pool. The per-item lift then uses the same formula as the single-stage setting:
\begin{equation*}
\sum_{k=1}^{K} w_{k}\,(\text{SNIPS}_{\text{Exploration},k} - \text{SNIPS}_{\text{Leave-One-Out},k}).
\label{eq:pool-incrementality}
\end{equation*}
Equivalently: if the upstream model had dropped $i$ from its pool, what user-engagement loss would follow? Aggregated across items, high lifts indicate the upstream model contributes genuine value, while consistently low lifts mean its pools are largely interchangeable downstream.

\subsubsection{Irreplaceability Estimators}
From user impression and engagement logs, we construct synthetic treatment samples for content item $i$: the treatment indicator $T_j$ denotes whether $i$ appeared in sample (e.g., session) $j$, and $r_j$ is the observed reward. Using the logged probability $\pi_j$ that $i$ was selected in impression $j$, we define $\text{Irreplaceability}(i)$, the Horvitz--Thompson~\cite{horvitz1952generalization} estimate over $N$ samples, to be
\begin{equation}
\frac{1}{N}\sum_{j=1}^{N}\left(\frac{T_j\, r_j}{\pi_j} - \frac{(1-T_j)\, r_j}{1-\pi_j}\right),
\label{eq:ht}
\end{equation}
and we define $\text{Irreplaceability}_{\text{H\'ajek}}(i)$, the H\'ajek estimate~\cite{hajek1971comment,swaminathan2015counterfactual}, to be 
\begin{align*}
\frac{\sum_{j\in\mathcal{E}_i} T_j\, r_j/\pi_j}{\sum_{j\in\mathcal{E}_i} T_j/\pi_j}
-
\frac{\sum_{j\in\mathcal{E}_i} (1-T_j)\, r_j/(1-\pi_j)}{\sum_{j\in\mathcal{E}_i} (1-T_j)/(1-\pi_j)},
\label{eq:hajek}
\end{align*}
where $\mathcal{E}_i$ is the set of samples in which $i$ is eligible. Horvitz--Thompson preserves unbiasedness under the usual assumptions but has high variance when propensities are uneven; H\'ajek is more stable in finite samples with reduced variance, at the cost of some bias, since it normalizes the inverse-propensity weights separately within the treated and control groups. We choose between them per observability scenario.

\subsubsection{Sequential Generation Without Replacement}
Many recommenders generate a sequence of content from a candidate pool (slates, playlists, feeds), selecting without replacement: once an item is chosen at an early step it is removed from the pool. We replace the per-sample propensity with a cumulative selection probability. If $\pi_{ik}$ is the logged probability that $i$ is selected at position $k$, conditional on not being selected at positions $1$ through $k-1$, then over a sequence of length $K$,
\begin{equation*}
p_{iK} = 1 - \prod_{k=1}^{K} (1 - \pi_{ik}),
\label{eq:page-probability}
\end{equation*}
and $p_{iK}$ replaces $\pi_j$ in Eq.~\ref{eq:ht} for sequential generation.

\subsubsection{Universality}
Universality is the fraction of samples (e.g., sessions) in which content $i$ is an eligible candidate:
\begin{equation*}
\text{Universality}(i) = \frac{\#\{\text{samples where content } i \text{ is a candidate}\}}{\#\{\text{all samples}\}}.
\label{eq:universality}
\end{equation*}

\subsection{Simulations Setup}
The simulation is set up as follows.

\subsubsection{Individual Content}
In our simulation, the arm space is a collection of 1,000 individual pieces of content, each characterized by an intrinsic CTR sampled from a $Beta(1, 8)/4$ distribution. The distribution is chosen to reflect a skewed distribution where the majority of content exhibits lower engagement probabilities, while a few demonstrate much higher user affinity.

\subsubsection{Content Set}
A content set is defined as a structured list of content, subject to a constraint defined by a viewing depth (i.e., 10 in this setup). We simulate the creation of two types of policies:
\begin{itemize}
\item Target policy: The content recommender has access to a noisy realization of user content affinity (i.e., user content affinity + $N(0, noise)$) and ranks the content according to its perceived affinity. The noise level is an indicator of the quality of recommendation model.
\item Explore policy or logging policy: The content recommender selects the content to recommend with a certain level of randomness. For simplicity, here we use a Boltzmann distribution with a temperature of 0.1. During this process, the selection probabilities of each content candidate are logged.
\end{itemize}

\subsubsection{User}
Each user exhibits unique affinities towards each content in the content set. These affinities are modeled as noisy realizations of the inherent content CTR, with the noise following a normal distribution where the standard deviation is CTR. The affinities are then linearly scaled into the range [min(inherent\_CTR), max(inherent\_CTR)].

When presented with a set of recommended content, a user decides whether to click by following this decision process: The viewing depth is first sampled from a $Poisson(10)$ distribution. The user then views content sequentially until they reach this view depth or the end of available content. For each content, the CTR follows a $Bernoulli(\text{affinity}_i)$ distribution. If any content within the view depth results in a positive realization, the content set is considered ``clicked''; otherwise, the content set is ``abandoned.''

\end{document}